\documentclass[aps,prl,twocolumn,groupedaddress]{revtex4-2}

\usepackage{bm,amssymb,graphicx,hyperref,color}
\usepackage{amsmath}

\newcommand{\bs}[1]{\boldsymbol{#1}}
\newcommand{\bsr}{\bs{r}}
\newcommand{\bsv}{\bs{v}}
\newcommand{\bsu}{\bs{u}}
\newcommand{\bsc}{\bs{c}}

\begin{document}


\title{Classifying exact local-equilibrium solutions of the Boltzmann equation}


\author{Nagi Khalil}
\email[]{nagi.khalil@uah.es}
\affiliation{Departamento de Física y Matemáticas \& Grupo Interdisciplinar de Sistemas Complejos (GISC), Universidad de Alcalá, 28805 Alcalá de Henares, Spain}
\author{Jorge Bravo-Gadea}
\email[]{jorge.bravog@uah.es}
\affiliation{Departamento de Física y Matemáticas, Universidad de Alcalá, 28805 Alcalá de Henares, Spain}


\date{\today}

\begin{abstract}
  We provide a complete classification, in a \(d\)-dimensional Euclidean space, of all hydrodynamic states and external forces for which a local equilibrium distribution constitutes an exact solution to the Boltzmann kinetic equation. We demonstrate that the temperature \(T\) must be spatially uniform, that the velocity field is the sum of a rigid translation, a radial expansion or compression, and a rigid rotation, and that the external acceleration must be conservative, though time-dependent within a reference frame that translates and rotates along with the fluid. The density profile is then determined by the external potential, while its temporal dependence is governed by any temperature function \(T(t)\) that can be generated through an appropriate adjustment of the external acceleration. We recover the classic ``harmonic breather'' and obtain solutions involving static forces that exhibit algebraic cooling, similar to Haff's law, despite the occurrence of elastic collisions. Our classification identifies the exceptional circumstances under which local equilibrium is exact and provides rigorous criteria for evaluating kinetic and hydrodynamic approximations.
\end{abstract}


\maketitle

\textit{Introduction.---} The dynamics of a broad class of systems composed of distinguishable units/particles can be split into two separate contributions: a free evolution, akin to free streaming, punctuated by binary interactions. Kinetic Theory provides a suitable mathematical description for such systems in terms of the one-particle distribution function \(f(\bsr,\bsv,t)\), defined such that \(f(\bsr,\bsv,t)d\bsr d\bsv\) is the expected number of particles in the corresponding phase-space element. For monodisperse, classical particles in \(d\) dimensions, \(f\) satisfies
\begin{equation}
  \label{eq:kineticeq}
  \partial_t f+\bsv\cdot \partial_{\bsr} f+\bs g\cdot \partial_{\bsv}f=J,
\end{equation}
where \(\bs g(\bsr,t)\) is the acceleration of an isolated particle and \(J\) a collision operator. In general, the latter is a function of position and time, takes into account pre-collision correlations, and hence includes information beyond \(f\). However, when collision events are statistically independent, then the collision operator is a functional of \(f\) solely, and the equation reduces to the well-known Boltzmann kinetic equation (BE) \cite{dorobe21}. 

The BE has attracted significant interest within the physics and mathematics communities since the inception of Kinetic Theory by Maxwell and Boltzmann \cite{ma67,bo96}. Its study establishes a link between the microscopic properties of particles and the macroscopic laws observed in everyday life, in most cases involving the so-called hydrodynamic fields: number density \(n=\int d\bs vf\), velocity field \(\bs u=\frac{1}{n}\int d\bs v\bsv f\), and temperature \(T=\frac{1}{dn}\int (\bs v-\bs u)^2f\). From a mathematical perspective, considerable effort has been devoted to the sixth of Hilbert's problems: the derivation of fluid equations from Boltzmann's Kinetic Theory \cite{vi02,la05,sa09,gasate13,bogasasi23,dehama24,dehama25}.

However, even though the links between the different levels of description can be rigorously established, there still remains the problem of solving Eq.~\eqref{eq:kineticeq}. As a consequence, a common approach when describing a complex system is to consider a local equilibrium approximation: the distribution function of the system, while out of equilibrium, is approximated by
\begin{equation}
  \label{eq:scalingf}
  f_{{eq}}(\bsr,\bsv,t)=n(\bsr,t)[v_0(\bsr,t)]^{-d}\phi(\bsc)
\end{equation}
with \(\bsc=\frac{\bsv-\bsu}{v_0}\) a dimensionless peculiar velocity, \(v_0=\sqrt{2T}\) a thermal velocity, and \(\phi=\pi^{-\frac{d}{2}}e^{-c^2}\) the Gaussian distribution with zero mean and variance equal to \(d/2\). This ansatz replaces the constant parameters of a global Maxwellian by local hydrodynamic fields \cite{sapubu25}. However, a basic question remains largely unanswered: for which states and external forcings is the local-equilibrium ansatz an exact, rather than approximate, solution?

There are already relevant studies identifying local equilibrium solutions for the BE beyond the global equilibrium solution; Boltzmann himself identified one for a system subject to an external harmonic potential \cite{bo09}; see also the works of Grad \cite{gr49,gasa03}. Recent works have provided more examples for systems subject to an external conservative force, eventually time-dependent \cite{scla09,gumurutr14,lobacole15,gujaplprtr23,gamagutr24,magagutr24,cadohemimosc24} with eventual rotations \cite{kiyu12}. In this letter we go one step further and provide a complete classification \(d\)-dimensional Euclidean space: we state what kind of external acceleration is compatible with local equilibrium and how the hydrodynamic fields depend on space and time. 

Although boundary and initial conditions play an important role in solving kinetic equations, see for example the formal approach in references \cite{brde20,do26}, they do not play a fundamental role in our reasoning. We simply identify all possible local equilibrium solutions, without regard for the necessary conditions.

\textit{Hydrodynamic problem.---} A local Maxwellian is completely determined by its density, flow velocity, and temperature. We therefore seek the conditions that these fields must satisfy for the Maxwellian to solve the kinetic equation exactly. For that purpose it is enough, as we shall see, that the collision operator in \eqref{eq:kineticeq} vanishes when the distribution function of the system is given by Eq.~\eqref{eq:scalingf}:
\begin{equation}
  J[f_{eq}]=0.
\end{equation}
This holds exactly for most operators that conserve particle number, linear momentum, and energy, but it may also be valid beyond this case. In particular, it is exact for the elastic Boltzmann collision operator, under the standard microreversibility assumptions, with any inter-particle interaction potential. Furthermore, it can serve as a good approximation beyond the dilute limit, through the use of Enskog-type operators \cite{gamagutr24,magagutr24}.

We now proceed by deriving a set of hydrodynamic equations that a system in local equilibrium must verify. The details of the computation can be found in the Supplemental Material. Using the scaling \eqref{eq:scalingf} with Eq.~\eqref{eq:kineticeq}, collecting independent monomials in the peculiar velocity, we obtain
\begin{equation}
  \label{eq:alpha+betai}
  \alpha +\beta_i c_i+ \gamma c^2+\eta_{ij}c_ic_j+\theta_ic^2c_i=0,
\end{equation}
where \(\alpha\), \(\beta_i\), \(\gamma\), \(\eta_{ij}\), and \(\theta_i\) are functions of \(\bsr\) and \(t\) through the hydrodynamic fields \cite{seesm}. Because this polynomial identity holds for every \(\bs c\), each independent tensorial coefficient must vanish, implying that \(\alpha=\beta_i=\theta_i=0\), \(\gamma+\frac{1}{d}\eta_{ii}=0\) and \(\eta_{ij}-\frac{\eta_{kk}}{d}\delta_{ij}\) must be antisymmetric. These conditions give rise to the following hydrodynamic relations \cite{seesm}
\begin{align}
  \label{eq:siseq1}
  & \partial_{\bsr} T=\bs 0, \\
  \label{eq:siseq2}
  & \partial_{r_i}u_j=\epsilon \delta_{ij}+w_{ij}, \\
  \label{eq:siseq3}
  & \partial_t n+\partial_{\bsr}\cdot(n\bsu)=0, \\
  \label{eq:siseq4}
  & \partial_t \bsu+\bsu\cdot \partial_{\bsr}\bsu+T\partial_{\bsr}\ln n-\bs g=\bs 0, \\
  \label{eq:siseq5}
  & \partial_t \ln T+2\epsilon=0,
\end{align}
where \(\epsilon=\frac{1}{d}\partial_{\bsr}\cdot \bsu\) and \(w_{ij}=\frac{1}{2}(\partial_{r_i}u_j-\partial_{r_j}u_i)\).

\textit{Spatial dependence.---} From Eq.~\eqref{eq:siseq1} we see that the temperature is spatially homogeneous. Hence, in general, it can depend only on time
\begin{equation}
  T=T(t).
\end{equation}
This implies, from Eq.~\eqref{eq:siseq5}, that the divergence of the velocity field \(\epsilon\) can only depend on time as well. Moreover, it can also be seen that the term \(w_{ij}\) in Eq.~\eqref{eq:siseq2} is also spatially homogeneous \cite{seesm}. With these results in mind, Eq.~\eqref{eq:siseq3} can be integrated to obtain the general spatial form of the velocity field:
\begin{equation}
  \label{eq:uiui0etwji}
  u_i(\bs r,t)=u_i^{(0)}(t)+\epsilon(t) r_i+ w_{ji}(t)r_j
\end{equation}
where \(\bs u^{(0)}\) is an arbitrary time-dependent vector. We observe three different contributions: \(u_i^{(0)}\) represents a rigid translation,  \(\epsilon(t) r_i\) is a radial contribution, and \(w_{ji}(t)r_j\), due to the antisymmetric character of \(w_{ij}\), gives rise to a rotation.

In order to simplify the subsequent analysis, it is convenient to move from our current inertial frame of reference \(\mathcal F\) to a new frame \(\mathcal F^*\), one that translates with velocity \(u_i\) and rotates according to \(w_{ij}\). This is possible because these quantities are position independent. It can be seen \cite{seesm} that in the new frame \(\mathcal F^*\) the system \eqref{eq:siseq1}--\eqref{eq:siseq5} retains the same form except that Eq.~\eqref{eq:siseq2} loses the term \(w_{ij}\) and Eq.~\eqref{eq:siseq5} acquires additional non-inertial contributions. Therefore, in the new frame the hydrodynamics is governed by the system \eqref{eq:siseq1}--\eqref{eq:siseq5} with \(w_{ij}=0\) and \(\bs g\) absorbing the additional non-inertial terms. This way, the velocity field in \(\mathcal F^*\) is purely radial:
\begin{equation}
  \label{eq:bsuet}
  \bs u=\epsilon(t)\bsr.
\end{equation}

From now on, we work in \(\mathcal F^*\) and, to simplify the exposition, we keep the same notation. Plugging Eq.~\eqref{eq:bsuet} into Eq.~\eqref{eq:siseq4}, we have
\begin{equation}
  \label{eq:eqbsrlnn2}
  \partial_{\bsr}\left(\ln n+\frac{\epsilon^\prime+\epsilon^2}{2T}r^2\right)=\frac{1}{T}\bs g,
\end{equation}
where \(\mbox{}^\prime\) indicates time derivative. The previous relation implies that \(\bs g\) must be a conservative field: there exists a scalar field \(E_g(\bsr,t)\) such that
\begin{equation}
  \bs g (\bsr,t)=-\partial_{\bsr} E_g(\bsr,t).
\end{equation}
In terms of the new scalar function \(E_g\), that depends on time in general, we can integrate Eq.~\eqref{eq:eqbsrlnn2} to have the density field 
\begin{equation}
  \label{eq:nbsrt}
  n(\bsr,t)=n_0(t)e^{-a(t)r^2-E_g(\bsr,t)/T(t)},
\end{equation}
with \(n_0(t)\) an integration function, arbitrary function of time, and \(a=\frac{\epsilon^\prime+\epsilon^2}{2T}\) that, recalling Eq.~\eqref{eq:siseq5}, is a function of \(t\) only, through \(T(t)\).

The fact that \(\bs g\) is conservative in the rotating reference frame \(\mathcal F^*\) implies that the origin of the time-dependent rotations in the velocity field, expressed in Eq.~\eqref{eq:uiui0etwji} for the original reference frame \(\mathcal{F}_1\), stems from the non-conservative component of the original acceleration. Furthermore, this non-conservative component cannot be arbitrary; rather, it is contained within the non-inertial terms that arise when transitioning from frame \(\mathcal F^*\) to \(\mathcal F\). In particular, if the acceleration is conservative in the original reference frame, then \(w_{ij}\) can only be constant.

\textit{Time dependence.---} Now we come to the time dependence of the hydrodynamic fields. Using Eqs.~\eqref{eq:nbsrt} and \eqref{eq:bsuet} for \(n\) and \(\bs u\) in Eq.~\eqref{eq:siseq3}, we have
\begin{equation}
  \label{eq:a+2er2}
  -(a^\prime+2a\epsilon)r^2+\frac{\epsilon}{T}\bs g\cdot \bsr-\partial_t\left(\frac{E_g}{T}\right)+d\epsilon+\partial_t\ln n_0=0.
\end{equation}
In order to analyze this equation, it is useful to distinguish between two cases: \(T^\prime=0\) and \(T^\prime\ne 0\).

When \(T^\prime=0\), spatial homogeneity makes \(T\) uniform and constant, \(T=T_0\ge 0\) with \(T_0=T(0)\). Moreover, since \(T^\prime=0\) is equivalent to \(\epsilon=0\), using \(\epsilon=0\) in Eq.~\eqref{eq:bsuet}, the velocity field vanishes \(\bs u=\bs 0\). As for the number density, using \(T^\prime=\epsilon=0\) in Eq.~\eqref{eq:a+2er2}, we obtain \(-\partial_t\left(\frac{E_g}{T}\right)+\partial_t\ln n_0=0\) implying \(\partial_t\ln \left(n_0e^{-E_g/T}\right)=0\). Hence, the density reads \(n(\bsr)=n_0e^{-E_g(\bsr)/T_0}\) with \(n_0\) a nonnegative constant. In conclusion, when \(T=T_0\), or equivalently \(\epsilon=0\), we have global equilibrium. Moreover, the external conservative acceleration has to be time independent. Strictly speaking, when \(E_g\) is uniform, we could choose \(E_g\) to depend on time if \(n_0\) also depends on time in a way that \(n_0e^{-E_g/T_0}\) is constant; however this \(E_g(t)\) is physically irrelevant: it does not produce any acceleration. 

In the case \(T^\prime \ne 0\), assuming \(T^\prime\) vanishes at most on a discrete set, we can define a new time variable \(\tau \) on any time interval where \(T^\prime\ne 0\) by \(\tau=\int_0^t\epsilon(t^*)dt^*=\frac{1}{2}\ln\left(\frac{T_0}{T(t)}\right)\). Using \(\tau\), Eq.~\eqref{eq:a+2er2} can be written as
\begin{equation}
  \label{eq:mathcalOEgT}
  \mathcal O\left(\frac{E_g}{T}\right)=-(\dot a+2a)r^2+d+\partial_\tau\ln n_0,
\end{equation}
where \(\dot a=\frac{da}{d\tau}\) and \(\mathcal O=r\partial_r+\partial_\tau\). The general solution to this equation is \cite{seesm}
\begin{equation}
  \label{eq:egttehbis}
  E_g(\bsr,t)=T(t)V(r\sqrt{T(t)/T_0},\hat{\bsr})-a(t)T(t)r^2,
\end{equation}
where \(V(r,\hat{\bsr})\) is a generic function of  \(r\), the modulus of \(\bsr\), and \(\hat{\bsr}\equiv(\theta_1,\dots,\theta_{d-1})\), all angles of the \(d\)-dimensional spherical coordinates. In this case, the density and energy fields can be written as
\begin{align}
  \label{eq:bsun0}
  & n(\bsr,t)=n_0\left[\frac{T(t)}{T_0}\right]^{\frac{d}{2}}\exp\left[-V(r\sqrt{T(t)/T_0},\hat{\bsr})\right], \\
  \label{eq:bsuTprime}
  & \bsu=-\frac{T^\prime(t)}{2T(t)}\bs r,   
\end{align}
with \(n_0\) and \(T_0\) non-negative constants. Notice that for \(T=T_0\), even though the procedure is not applicable, the solution of global equilibrium is recovered. Hence, at this level, the actual time evolution of \(T(t)\) is rather generic, provided the external acceleration is chosen properly according to Eq.~\eqref{eq:egttehbis}. If, instead, the external acceleration is chosen with a given form, then a state of local equilibrium may not exist, may exist for a specific form of  \(T(t)\), or for more than one possible function \(T(t)\).

As said, the case \(T^\prime=0\) corresponds to global equilibrium: constant, generically non-uniform density and zero velocity field. We obtain a sort of reciprocal result. If \(\bs u=0\) then, from Eq.~\eqref{eq:bsuTprime}, it is also \(T^\prime=0\) and we also have global equilibrium. However, we can have a steady-state density profile at local equilibrium without being global. It is enough to choose \(V=-d\ln\left(\sqrt{\frac{T}{T_0}}r\right)\) with \(T^\prime\ne 0\), then \(n=n_0r^{-d}\) is time independent.

\textit{Symmetries.---} Looking at Eq.~\eqref{eq:egttehbis} we identify two terms that may have different symmetries. On the one hand, due to its dependence on \(\hat{\bsr}\), \(V\) may not be spherically symmetric. On the other hand, the second term is clearly isotropic. Interestingly, the two terms affect differently the hydrodynamic fields. As it is apparent from Eqs.~\eqref{eq:bsun0} and \eqref{eq:bsuTprime}, the potential directly affects the number density \(n\) only through \(V\) while the velocity field remains always radial.

\begin{figure}
 \includegraphics[width=0.95\linewidth]{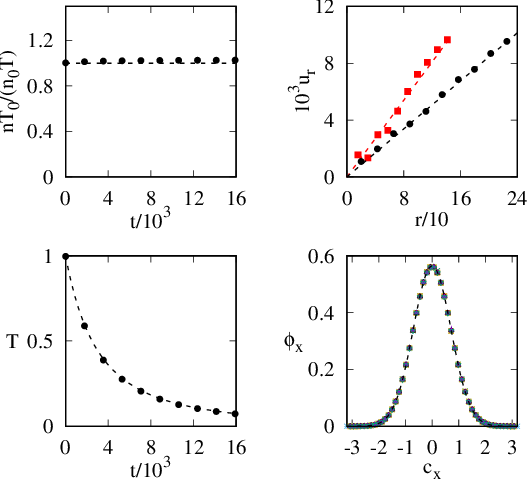}%
 \caption{Event-driven numerical simulations (symbols) versus theory (dashed lines) for a system of hard disks inside a circular system expanding adiabatically at constant velocity \cite{seesm}. The plots show: (a) \(nT_0/(n_0T)\) as a function of time; (b) the radial component velocity field \(u_r\) versus the distance to the center of the system for two values of the time: an intermediate one (red squares) and the final one (black circles); (c) the temperature \(T\) as a function of time; and (d) the marginal scaling distribution function \(\phi_x(c_x)\) along the \(X\)-direction measured at the final time for different distances to the origin.}
 \label{fig:fig1}
\end{figure}

Another interesting feature of Eq.~\eqref{eq:egttehbis} concerns its time dependence. It could be possible to have time independent acceleration, with \(E_g=E_g(\bsr)\), even with time dependent temperature \(T(t)\). Beyond the case \(E_g=0\) that will be analyzed as an example below, this situation can be identified by requiring both terms in Eq.~\eqref{eq:egttehbis} to be time independent separately. For the term involving \(V\), the most general form of \(V\) that makes the first term time independent is
\begin{equation}
  V(r,\hat{\bsr})=\frac{\varphi(\hat{\bsr})}{r^2},
\end{equation}
with \(\varphi(\hat{\bsr})\) an arbitrary function. The simplest case of \(V=0\) corresponds to choosing \(\varphi=0\). For the second condition, we need \(a(t)T(t)\) to be constant. Choosing for simplicity \(a(t)T(t)=0\), we have \(\frac{2}{T}\left(\frac{1}{T}\right)^{\prime\prime}-\left[\left(\frac{1}{T}\right)^\prime\right]^2=0\) and 
\begin{equation}
  \label{eq:thaff}
  T(t)=\frac{T_0}{\left(\frac{T_0^\prime}{T_0}t-\frac{1}{2}\right)^2+\frac{3}{4}},
\end{equation}
where \(T_0^\prime\) coincides with \(T'(0)\) and is an integration constant. This has the form of Haff's law for granular gases \cite{ha83,kh18}. However, here the source of ``cooling'' is not the dissipation in collisions but adiabatic expansion/compression. As a result, we can have cooling for \(T_0^\prime<0\), heating and then cooling for \(T_0^\prime >0\), or global equilibrium for \(T_0^\prime=0\). We illustrate the first case in Fig.~\ref{fig:fig1} where we show event-driven numerical simulation results, along with the theoretical results given by Eqs.~\eqref{eq:bsun0}, \eqref{eq:bsuTprime}, \eqref{eq:thaff}, for the simplest case \(\varphi=0\). 

\textit{Examples.---} Let us explore some illustrative examples of the external acceleration. The simplest case corresponds to \(\bs g=\bs 0\) for which we can choose \(E_g=0\). As a consequence, from Eq.~\eqref{eq:egttehbis}, we have \(T(t)V(r\sqrt{T(t)/T_0},\hat{\bsr})-a(t)T(t)r^2=0\). As expected, \(T(t)=T_0\) is recovered by choosing \(V=0\). However, we can choose a function \(V\) so that \(T\) is not constant \cite{seesm}. Interestingly, the most general \(T(t)\) compatible with this potential decays as \(t^{-2}\) for large \(t\), again as Haff's law. We have a similar result for a homogeneous external acceleration \(\bs g(t)\) provided \(\bs g(t)/[T(t)]^{3/2}\) is a constant vector \cite{seesm}. In the present case, the spherical symmetry of the velocity field coexists with the translational symmetry of the external acceleration. Another interesting family of potentials are of the form \(\bs g=g_n(t)r^n\bsr\) for \(n\in\mathbb Z\). For \(n\ne -2,0\) it can be seen that \(g_n(t)/[T(t)]^{\frac{n+4}{2}}\) should be constant \cite{seesm}. For \(n=-2\), \(g_{-2}\) should be constant and the unique state is of global equilibrium. On the contrary, for \(n=0\), \(g_0(t)\) can be any function \cite{seesm}. 

\textit{Conclusions.---} We have identified all possible solutions to the BE that correspond to local equilibrium for which the distribution function scales with the hydrodynamic fields as in Eq.~\eqref{eq:scalingf}. When the external acceleration is conservative and time dependent, the system has a uniform temperature that depends on time in a way that can be completely adjusted by the external acceleration through Eq.~\eqref{eq:egttehbis}, the velocity field is always radial with the radial strain rate given by \(T(t)\), and the density field nonuniform and time dependent, in general. There is no need for the acceleration to depend on time to have a time dependent temperature, as we have identified in an example. Our classification provides a complete dictionary between external forcing and exact local-equilibrium hydrodynamics, and shows that Haff-like cooling can arise purely from the external field, without any dissipative collisions.





\section*{Acknowledgments}
Financial support is acknowledged through grant PID2023-151960NA-I00 funded by MCIN/AEI/10.13039/501100011033 and by ``ERDF A way of making Europe''.

\appendix

\bibliography{eqkt}

\end{document}


\title{Classifying exact local-equilibrium solutions of the Boltzmann equation \\
  \texttt{Supplemental Material}}

\author{Nagi Khalil}
\email[]{nagi.khalil@uah.es}
\affiliation{Departamento de Física y Matemáticas \& Grupo Interdisciplinar de Sistemas Complejos (GISC), Universidad de Alcalá, 28805 Alcalá de Henares, Spain}
\author{Jorge Bravo-Gadea}
\email[]{jorge.bravog@uah.es}
\affiliation{Departamento de Física y Matemáticas, Universidad de Alcalá, 28805 Alcalá de Henares, Spain}


\maketitle

\tableofcontents

\newpage

\section{Hydrodynamic problem}
\subsection{Derivation of Eq.~(4)}
\label{sec:hydroEqs}

Using the scaling \eqref{m-eq:scalingf} with the left-hand side of Eq.~\eqref{m-eq:kineticeq}, we have the following terms: 
\begin{align}
  \partial_t f&=nv_0^{-d}\phi(\bs c)\left[\partial_t\ln \left(nT^{-\frac{d}{2}}\right)+c_i\frac{2}{v_0}\cdot\partial_t u_i+c^2\partial_t \ln T\right], \\
  \bs v\cdot \partial_{\bs r} f&= nv_0^{-d}\phi(\bs c)\left(v_0\bsc+\bsu\right)\cdot\left[\partial_{\bsr}\ln \left(nT^{-\frac{d}{2}}\right)+c_i\frac{2}{v_0}\partial_{\bsr} u_i+c^2\partial_{\bsr} \ln T\right], \\
  \bs g\cdot \partial_{\bs v}f&=nv_0^{-d} \phi(\bsc)\left(-2v_0^{-1}c_i g_i\right).
\end{align}
With these relations Eq.~\eqref{m-eq:kineticeq}, with \(J[f_{eq}]=0\), gives rise to Eq.~\eqref{m-eq:alpha+betai} with
\begin{align}
  & \alpha(\bsr,t)=\frac{1}{nv_0}\left(\partial_t+\bsu\cdot\partial_{\bsr}\right)\ln \left(nT^{-\frac{d}{2}}\right),\\
  & \bs\beta(\bsr,t)=\frac{1}{n}\partial_{\bsr}\ln \left(nT^{-\frac{d}{2}}\right)+\frac{1}{nT}\left[\left(\partial_t+\bsu\cdot\partial_{\bsr}\right)\bsu-\bs g\right],\\
  & \gamma(\bsr,t)=\frac{1}{nv_0}\left(\partial_t+\bsu\cdot\partial_{\bsr}\right)\ln T, \\
  & \eta_{ij}(\bsr,t)=\frac{2}{nv_0}\partial_{r_i}u_j, \\
  &\bs\theta(\bsr,t)=\frac{1}{n}\partial_{\bsr}\ln T.
\end{align}

\subsection{Position independence of  \(w_{ij}\)}

Using that \(\epsilon=\epsilon(t)\) and, from Eq.~\eqref{m-eq:siseq2} and the asymmetry of \(w_{ij}\), the equality \(\epsilon\delta_{ij}=\frac{1}{2}(\partial_{r_i}u_j+\partial_{r_j}u_i)\), it follows that \(w_{ij}\) is position independent:
\begin{align}
  \partial_{r_k}w_{ij}&=\frac{1}{2}(\partial_{r_k}\partial_{r_i}u_j-\partial_{r_k}\partial_{r_j}u_i)=\frac{1}{2}(\partial_{r_i}\partial_{r_k}u_j-\partial_{r_j}\partial_{r_k}u_i) \nonumber \\ &= \frac{1}{2}[\partial_{r_i}(\partial_{r_k}u_j+\partial_{r_j}u_k)-\partial_{r_i}\partial_{r_j}u_k-\partial_{r_j}\partial_{r_k}u_i] \nonumber \\ &= \frac{1}{2}[\partial_{r_i}(\partial_{r_k}u_j+\partial_{r_j}u_k)-\partial_{r_j}(\partial_{r_i}u_k+\partial_{r_k}u_i)] \nonumber  \\ &= \partial_{r_i}\epsilon\delta_{jk}-\partial_{r_j}\epsilon\delta_{ik}=0 \nonumber  \\ & \Rightarrow  w_{ij}=w_{ij}(t).
\end{align}

\newpage

\section{Spatial and time dependence}
\subsection{Change of frame of reference}
\label{appen:1}

Let us change from the original frame of reference \(\mathcal F\) to a new one \(\mathcal F^*\) that translates and rotates. If \(r_i\) denotes the position coordinates of a point \(P\) in the old frame and \(r_i^*\) are the coordinates in the new frame, then 
\begin{equation}
  \label{m-eq:ri=ri0}
  r_i=r_i^{(0)}+(e^{\hat \theta})_{ij} r_j^*,
\end{equation}
where \(r_i^{(0)}\) accounts for the translation and \(\hat \theta\) it the (antisymmetric) matrix that generates the rotation; both are assumed to be functions of time. If \(P\) is moving, the relations between the velocities \(v_i\) and \(v_i^*\) are obtained by deriving the previous expression with respect to time:
\begin{equation}
  \label{m-eq:vivi0}
  v_i=v_i^{(0)}+(\hat \omega \cdot e^{\hat \theta})_{ij} r_j^*+(e^{\hat \theta})_{ij} v_j^*=v_i^{(0)}+\omega_{ij}(r_j-r^{(0)}_j)+(e^{\hat \theta})_{ij} v_j^*,
\end{equation}
where we have used Eq.~\eqref{m-eq:ri=ri0} in the last equality and introduced the following definitions: 
\begin{align}
  v_i^{(0)}&=\frac{d}{dt}r_i^{(0)}, \\
  \hat \omega &=\frac{d}{dt}\hat \theta.
\end{align}

Now we focus on the transformation of system \eqref{m-eq:siseq1}--\eqref{m-eq:siseq5}. In doing so, we have to change from the coordinates \((r_1,\dots,r_d,t)\) to \((r_1^*,\dots,r_d^*,t^*)\), where \(r_i(r_1^*,\dots,r_d^*,t^*)\) is given by Eq.~\eqref{m-eq:ri=ri0} and \(t(r_1^*,\dots,r_d^*,t^*)=t^*\). From these relations, we have the following useful equalities:
\begin{align}
  & \partial_{t^*}=(\partial_{t^*}r_i)\partial_{r_i}+(\partial_{t^*}t)\partial_t=[v_i^{(0)}+(\hat \omega\cdot e^{\hat \theta})_{ij}r_j^*]\partial_{r_i}+\partial_t, \\
  & \partial_{r_i^*}=(\partial_{r_i^*}r_j)\partial_{r_j}+(\partial_{r_i^*}t\partial_t)=(e^{\hat \theta})_{ji}\partial_{r_j}=(e^{-\hat \theta})_{ij}\partial_{r_j}.
\end{align}
The last equality holds because \(e^{\hat\theta}\) is orthogonal. 

Using Eq.~\eqref{m-eq:siseq1}, we have
\begin{equation}
  \partial_{r^*_i} T=-(e^{\hat\theta})_{ij}\partial_{r_j}T \quad \Rightarrow \quad \partial_{r^*_i} T=0.
\end{equation}
Proceeding similarly with Eq.~\eqref{m-eq:siseq5}:
\begin{equation}
  \partial_{t^*} \ln T=\partial_{t}\ln T \quad \Rightarrow \quad \partial_{t^*} \ln T+2\epsilon=0,
\end{equation}
where \(\epsilon\) can be considered as a function of \(t^*\).

As for the other equations, we have to include the correct transformation of the velocity field, from \(u_i\) to \(u_i^*\). Using \eqref{m-eq:vivi0}, the transformation reads
\begin{equation}
  \label{m-eq:uivi0}
  u_i=v_i^{(0)}+(\hat \omega \cdot e^{\hat \theta})_{ij}r_j^*+(e^{\hat \theta})_{ij} u_j^*=v_i^{(0)}+\omega_{ij}(r_j-r_j^{(0)})+(e^{\hat \theta})_{ij} u_j^*.
\end{equation}
This way, by deriving with respect to \(r_i^*\), we have
\begin{equation}
  \label{m-eq:ri*ujomegajk}
  \partial_{r_i^*}u_j=(\hat \omega\cdot e^{\hat \theta})_{ji}+(e^{\hat \theta})_{jk} \partial_{r_i^*}u_k^*.
\end{equation}
On the other hand, using Eq.~\eqref{m-eq:siseq2}, we also have
\begin{equation}
  \label{m-eq:pari*ujparkuj}
  \partial_{r_i^*}u_j=(e^{\hat \theta})_{ki}\partial_{r_k}u_j=(e^{\hat \theta})_{ki}(e\delta_{kj}+w_{kj}).
\end{equation}
Equating the previous two equations and, after some manipulations:
\begin{equation}
  (e^{\hat \theta})_{jk} \partial_{r_i^*}u_k^*=(e^{\hat \theta})_{ji}e+(e^{\hat \theta})_{ki}(w_{kj}-\omega_{jk}).
\end{equation}
Choosing
\begin{equation}
  \omega_{jk}=w_{kj}
\end{equation}
we finally have
\begin{equation}
  \partial_{r_i^*}u_j^*=\epsilon\delta_{ij}.
\end{equation}

For the transformation of Eq.~\eqref{m-eq:siseq3}, the following terms are useful:
\begin{align}
  \label{m-eq:partt*n}
  \partial_{t^*}n&=[v_i^{(0)}+(\hat\omega\cdot e^{\hat\theta})_{ij}r_j^*]\partial_{r_i}n+\partial_tn, \\
  \partial_{r_i^*}(n u_j)&=(e^{\hat \theta})_{ki}\partial_{r_k}(n u_j), \\
  \partial_{r_i^*}(n u_j)&=\partial_{r_i^*}\{n[v_j^{(0)}+(\hat \omega \cdot e^{\hat \theta})_{jk}r_k^*+(e^{\hat \theta})_{jk} u_k^*]\}\nonumber \\ &= [v_j^{(0)}+(\hat \omega \cdot e^{\hat \theta})_{jk}r_k^*](e^{\hat \theta})_{li}\partial_{r_l}n+(e^{\hat \theta})_{jk}\partial_{r_i^*}(nu_k^*).
\end{align}
Removing \(\partial_{r_i^*}(n u_j)\) from the last two equations:
\begin{align}
  \partial_{r_i^*}(nu_j^*)&=(e^{-\hat \theta})_{jl}(e^{\hat \theta})_{ki}\partial_{r_k}(n u_l) \nonumber \\ & -(e^{-\hat \theta})_{jm}[v_m^{(0)}+(\hat \omega \cdot e^{\hat \theta})_{mk}r_k^*](e^{\hat \theta})_{li}\partial_{r_l}n \\
  \label{m-eq:partrit*nui*}
  \Rightarrow \partial_{r_i^*}(nu_i^*) &= \partial_{r_i}(n u_i)-[v_i^{(0)}+(\hat \omega \cdot e^{\hat \theta})_{ij}r_j^*]\partial_{r_i}n.
\end{align}
Summing Eqs.~\eqref{m-eq:partt*n} and \eqref{m-eq:partrit*nui*}, and using Eq.~\eqref{m-eq:siseq3}, we obtain the equation in the new frame:
\begin{equation}
  \partial_{t^*}n+\partial_{r_i^*}(nu_i^*)=0.
\end{equation}

Finally, let us focus on Eq.~\eqref{m-eq:siseq4}. Proceeding as before, we have to consider different terms. Deriving Eq.~\eqref{m-eq:uivi0} with respect to \(t^*\):
\begin{equation}
  \label{m-eq:pt*uiai}
  \partial_{t^*}u_i=a_i^{(0)}+(\hat \omega^\prime \cdot e^{\hat \theta})_{ij} r_j^*+(\hat \omega^2\cdot e^{\hat \theta})_{ij} r_j^*+(\hat \omega \cdot e^{\hat \theta})_{ij} u_j^*+(e^{\hat \theta})_{ij}\partial_{t^*} u_j^*,
\end{equation}
where
\begin{equation}
  a_i^{(0)}=\frac{d}{dt}v_i^{(0)}
\end{equation}
and \({}^\prime\) denotes time derivative. Additionally:
\begin{equation}
  \label{m-eq:pt*uiprj}
  \partial_{t^*}u_i=[v_j^{(0)}+(\hat \omega\cdot e^{\hat \theta})_{jk} r_k^*]\partial_{r_j}u_i+\partial_tu_i.
\end{equation}
Equating Eqs.~\eqref{m-eq:pt*uiai} and \eqref{m-eq:pt*uiprj}:
\begin{align}
  \label{m-eq:part*uie}
  \partial_{t^*} u_i^*&=(e^{-\hat \theta})_{ij}\left\{\partial_{r_k}u_j\left[v_k^{(0)}+(\hat \omega\cdot e^{\hat \theta})_{kl} r_l^*\right]+\partial_tu_j \right. \nonumber \\ & \left. -\left[a_j^{(0)}+(\hat \omega^\prime \cdot e^{\hat \theta})_{jk} r_k^*+(\hat \omega^2\cdot e^{\hat \theta})_{jk} r_k^*+(\hat \omega \cdot e^{\hat \theta})_{jk} u_k^*\right] \right\}.
\end{align}
Using Eqs.~\eqref{m-eq:uivi0}, \eqref{m-eq:siseq4}, and then Eqs.~\eqref{m-eq:pari*ujparkuj} and \eqref{m-eq:ri*ujomegajk}
\begin{align}
  & \partial_{r_k}u_j\left[v_k^{(0)}+(\hat \omega\cdot e^{\hat \theta})_{kl} r_l^*\right]+\partial_tu_j= \partial_{r_k}u_j\left[u_k-(e^{\hat \theta})_{kl} u_l^*\right]+\partial_tu_j \nonumber \\ & =g_j-T\partial_{r_j}\ln n-(e^{\hat \theta})_{kl} u_l^*\partial_{r_k}u_j=g_j-T\partial_{r_j}\ln n-u_k^*\partial_{r_k^*}u_j \nonumber  \\ & = g_j-T\partial_{r_j}\ln n-u_k^*\left[(\hat \omega\cdot e^{\hat \theta})_{jk}+(e^{\hat \theta})_{jl} \partial_{r_k^*}u_l^*\right].
\end{align}
Combining this expression with Eq.~\eqref{m-eq:part*uie}, and the fact that 
\begin{equation}
  \partial_{r_j^*}\ln n=(e^{-\hat \theta})_{jk}\partial_{r_k}\ln n \quad  \Rightarrow \quad \partial_{r_j}\ln n=(e^{\hat \theta})_{jk}\partial_{r_k^*}\ln n,
\end{equation}
we finally have
\begin{align}
  & \partial_t^*u_i^*+u^*_j\cdot \partial_{r_j^*}u_i^*+T\partial_{r_i^*}\ln n-(e^{-\hat\theta})_{ij}g_j\nonumber \\ & \quad +(e^{-\hat\theta})_{ij}\left[a_j^{(0)}+(\hat \omega^\prime \cdot e^{\hat \theta})_{jk} r_k^*+(\hat \omega^2\cdot e^{\hat \theta})_{jk} r_k^*+2(\hat \omega \cdot e^{\hat \theta})_{jk} u_k^*\right]=0.
\end{align}

\subsection{Form of the external acceleration}

If we regard \(\frac{E_g}{T}\) as the unknown variable of Eq.~\eqref{m-eq:mathcalOEgT}, then by linearity:
\begin{equation}
  \frac{E_g}{T}=\frac{E_h}{T}+\frac{E_p}{T}
\end{equation}
where \(E_h/T\) is the general solution to the homogeneous equation 
\begin{equation}
  \label{m-eq:mathcalOEbsrt} 
  \mathcal O E(\bsr,t)=0
\end{equation}
and \(E_p/T\) is a particular solution of Eq.~\eqref{m-eq:mathcalOEgT}.

The general solution to Eq.~\eqref{m-eq:mathcalOEbsrt} is
\begin{equation}
  \frac{E_h(\bsr,t)}{T(t)}=V(re^{-\tau},\hat{\bsr})=V(r\sqrt{T(t)/T_0},\hat{\bsr}),
\end{equation}
where \(V\) is a generic function of \(d\) variables, differentiable with respect to its first argument, and \(\hat{\bsr}\) collects all angles of the \(d\)-dimensional spherical coordinates. A particular solution to Eq.~\eqref{m-eq:mathcalOEgT} is
\begin{equation}
  \frac{E_p(\bsr,t)}{T(t)}=-a(t)r^2+\ln\left(\frac{n_0(t)T^{-\frac{d}{2}}(t)}{n_0(0)T_0^{-\frac{d}{2}}}\right).
\end{equation}

With the previous results, a potential consistent with a generic time-dependent temperature \(T(t)\) is of the form
\begin{equation}
  \label{m-eq:egtteh}
  E_g(\bsr,t)=T(t)V(r\sqrt{T(t)/T_0},\hat{\bsr})-a(t)T(t)r^2+T(t)\ln\left[\frac{n_0(t)T^{-\frac{d}{2}}(t)}{n_0(0)T_0^{-\frac{d}{2}}}\right],
\end{equation}
where \(V\) is a generic function and \(a(t)\) is given by \(T(t)\):
\begin{align}
  \label{eq:afepsionlp}
  & a=\frac{\epsilon^\prime+\epsilon^2}{2T}, \\
  \label{eq:epftp2t}
  & \epsilon=-\frac{T^\prime}{2T}.
\end{align}

\newpage
\section{Examples}

\subsection{Uniform external acceleration}

A uniform external field \(\bs g(t)\) can be generated by \(E_g=-\bs g(t)\cdot \bs r\). Imposing Eq.~\eqref{m-eq:egttehbis}, we have
\begin{equation}
  \label{m-eq:bsgtcdotbsr}
  -\bs g(t)\cdot \bs r=T(t)V(r\sqrt{T(t)/T_0},\hat{\bsr})-a(t)T(t)r^2,
\end{equation}
which allows us to identify the most general forms of \(V\) and \(a\). Namely, since the left-hand side is proportional to \(r\), then \(V\) must be of the form
\begin{equation}
  V(r,\hat r)=k_1r^2+k_2\bs g_0\cdot \bs r,
\end{equation}
with \(k_1\) and \(k_2\) real constants and \(\bs g_0\) a constant vector. Inserting this expression of \(V\) in Eq.~\eqref{m-eq:bsgtcdotbsr}, we have
\begin{align}
  k_1\frac{T(t)}{T_0}&=a(t), \\
  k_2T(t)\sqrt{\frac{T(t)}{T_0}}\bs g_0\cdot \bs r&=-\bs g(t)\cdot \bsr.
\end{align}
Evaluating the first relation at \(t=0\), it is 
\begin{equation}
  k_1=a_0
\end{equation}
with \(a_0=a(0)\). In addition, by choosing \(k_1\) as
\begin{equation}
  k_2=-\frac{1}{\sqrt{T_0}},
\end{equation}
we infer the following relations:
\begin{align}
  \bs g(t) &=\left[\frac{T(t)}{T_0}\right]^{\frac{3}{2}}\bs g_0, \\
  \label{eq:ata0t0tt}
  a(t) &=\frac{a_0}{T_0}T(t).
\end{align}

An equation for \(T\) can now be obtained from Eq.~\eqref{eq:ata0t0tt}, using the definition of \(a\) in Eq.~\eqref{eq:afepsionlp} and Eq.~\eqref{eq:epftp2t}:
\begin{equation}
  \label{m-eq:casegg0vec}
  \frac{2}{T}\left(\frac{1}{T}\right)^{\prime\prime}-\left[\left(\frac{1}{T}\right)^\prime\right]^2=8\frac{a_0}{T_0}.
\end{equation}
Its general solution reads 
\begin{equation}
  \label{m-eq:tempcasoi}
  T(t)=T_0\left[\left(2a_0T_0+\frac{T_0^{\prime 2}}{T_0^2}\right)t^2-\frac{T_0^\prime}{T_0}t+1\right]^{-1}
\end{equation}
with \(T_0^\prime=T^\prime(0)\). By deriving Eq.~\eqref{m-eq:tempcasoi}, it can be seen that \(a_0=(-2T_0^{\prime\prime}/T_0+3{T_0^\prime}^2/T_0^2)/(8T_0)\), where \(T_0^{\prime\prime}=T^{\prime\prime}(0)\). Hence, \(T(t)\) is given by its initial value and that of its first two derivatives.

Finally, the other hydrodynamic fields are
\begin{align}
  \label{m-eq:nbsrteq}
  n(\bsr,t)&=n_0\left[\frac{T(t)}{T_0}\right]^{\frac{d}{2}}\exp\left\{-\frac{a_0T(t)}{2T_0}r^2+\frac{[T(t)]^{1/2}}{T_0^{3/2}}\bs g_0\cdot \bs r\right\}, \\
  \label{m-eq:ubsrteq}
  \bs u(\bs r,t)&=-\frac{T^\prime}{2T}\bsr=\frac{\left(2a_0T_0+\frac{T_0^{\prime 2}}{T_0^2}\right)t-\frac{T_0^\prime}{2T_0}}{\left(2a_0T_0+\frac{T_0^{\prime 2}}{T_0^2}\right)t^2-\frac{T_0^\prime}{T_0}t+1}\bs r.
\end{align}

\subsection{Case \(\boldsymbol{g}=g_n(t) r^n \boldsymbol{r}, \, n\in \mathbb Z-\{-2,0\}\)}

For this case, we can take \(E_g=-\frac{g_n(t)}{n+2}r^{n+2}=-\frac{1}{n+2}\bs g\cdot \bs r\). We have a case similar to the previous one. The most general form of \(V\) compatible with \(E_g\) and Eq.~\eqref{m-eq:egttehbis} is 
\begin{equation}
  V(r,\hat r)=k_1r^2+k_2r^{n+2}
\end{equation}
with
\begin{align}
  k_1\frac{T(t)}{T_0} &=a(t), \\
  k_2T(t)\left(\frac{T(t)}{T_0}\right)^{\frac{n+2}{2}} &=-\frac{1}{n+2}g_n(t).
\end{align}
Evaluating at \(t=0\):
\begin{align}
  k_1 &=a_0, \\
  k_2 &=-\frac{1}{n+2}\frac{g_n(0)}{T_0}.
\end{align}
This way, \(T(t)\) is also given by Eq.~\eqref{m-eq:tempcasoi} while the external acceleration has to be of the form 
\begin{equation}
  \bs g=g_n(0)\left[\frac{T(t)}{T_0}\right]^{\frac{n+4}{2}}r^n\bs r,
\end{equation}
with \(g_n(0)\) a constant. The density is given by Eq.~\eqref{m-eq:nbsrteq} after replacing the contribution \(\frac{[T(t)]^{1/2}}{T_0^{3/2}}\bs g_0\cdot \bs r\) in the exponential by \(\frac{[T(t)]^{(n+2)/2}}{T_0^{(n+4)/2}}g(0)\frac{r^{n+2}}{n+2}\) and the velocity field is given by Eq.~\eqref{m-eq:ubsrteq}. 

\subsection{Case \(\bs g=g_{-2}(t)r^{-2}\bs r\)}

Taking \(E_g=-g_{-2}(t)\ln r\), the most general function \(V\) is
\begin{equation}
  V(r,\hat r)=k_1r^2+k_2\ln r+k_3
\end{equation}
with \(k_3\) also a real constant. Proceeding as in the previous cases, we obtain
\begin{align}
  k_1\frac{T(t)}{T_0} &=a(t), \\
  k_2 &=-\frac{g_{-2}(t)}{T(t)}, \\
  k_3 &=-k_2\ln\frac{T(t)}{T_0}.
\end{align}
There are two possibilities. Analyzing the third equation:
\begin{itemize}
\item[-] The temperature is not constant: \(T(t)\ne T_0\), and \(k_2=k_3=0\). Then \(g_{-2}=0\).
\item[-] The temperature is constant: \(T=T_0\), then \(g_{-2}\) is also constant. This gives rise to global equilibrium. 
\end{itemize}

\subsection{Case \(\bs g=g_0(t)\bs r\)}

By choosing \(E_g(\bsr,t)=-\frac{1}{2}g_0(t)r^2\), from Eq.~\eqref{m-eq:egttehbis} the most general form of the function \(V\) is
\begin{equation}
  V(r,\hat r)=k_1r^2,
\end{equation}
with 
\begin{equation}
  k_1=a(t)-\frac{g_0(t)}{T(t)}.
\end{equation}
The constant \(k_1\) is
\begin{equation}
  k_1=a_0-\frac{g_0(0)}{T_0}.
\end{equation}
With these two previous expressions, and proceeding as before, the equation for \(T\) now reads
\begin{equation}
  \frac{2}{T}\left(\frac{1}{T}\right)^{\prime\prime}-\left[\left(\frac{1}{T}\right)^\prime\right]^2-\frac{4g_0(t)}{T^2}=\frac{8}{T_0}\left(a_0-\frac{g_0(0)}{2T_0}\right).
\end{equation}
This equation reduces to Eq.~\eqref{m-eq:casegg0vec} for \(g_0=0\), as expected, but also for \(g_0\) proportional to \(T^2\). There is no restriction on the functional form \(g_0(t)\).

\newpage
\section{Numerical simulations}

The numerical results presented in Fig.~\ref{m-fig:fig1} of the main text were obtained using an event-driven algorithm \cite{lu91}, employed to simulate a system in \(d=2\) consisting of \(200\) hard disks with diameter \(1\) and mass \(1\) (these values fix the units of length and mass, respectively). The particles are contained within a circle of radius \(R\), which increases at a constant velocity of \(0.01\) (this, together with the particle diameter, defines the unit of time), growing from an initial value of \(60\) to a final value of \(240\). Consequently, the simulation covers a time interval of \(18 \cdot 10^3\). The particles move freely between elastic collisions, occurring both among the particles themselves (conserving linear momentum) and with the circular boundary (which is assumed to have infinite mass and only alters the radial component of a colliding particle's velocity). 

The initial condition is that all particles are (approximately) in equilibrium at a temperature \(1\). This implies that they are distributed (approximately) uniformly, without overlaps, and exhibit a Gaussian velocity distribution. Throughout the temporal evolution of the system, we have measured the number density \(n(r_i,t_j)\), the radial and transverse velocity fields, \(u_r(r_i,t_j)\) and \(u_t(r_i,t_j)\), the temperature \(T(r_i,t_j)\), and the scaled marginal distribution function \(\phi_x(r_i)\). The values \(r_i\) and \(t_j\) correspond to \(10\) distinct annular strips, characterized by an average distance from the origin \(r_i\) and a width of \(R(t_j)/10\), evaluated at different time points \(t_j\), where \(t_j\) represents one of the \(10\) values obtained by dividing the total time interval. The marginal distribution was measured only at the final time point.

Measurements imply different averages:
\begin{itemize}
\item[-] Time: after each value \(t_j\), we have measure over \(10^3\) times in a narrow time window of width \(18\).
\item[-] Space: measurement are also averaged over the same strip.
\item[-] Realizations/trajectories: a total of \(10^5\). 
\end{itemize}

\begin{figure}[!h]
 \includegraphics[width=0.75\linewidth]{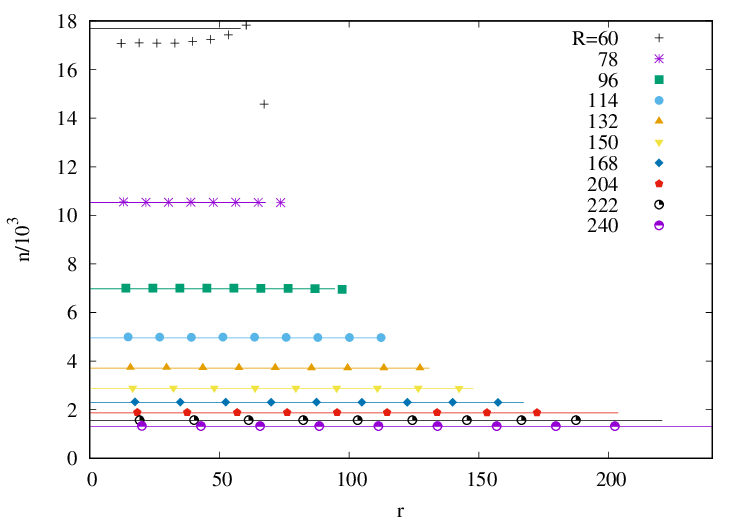}%
 \caption{Number density \(n\) as a function of \(r\), the distance to the origin, for different system radii \(R\).}
  \label{fig:fig2}
\end{figure}

\begin{figure}[!h]
 \includegraphics[width=0.75\linewidth]{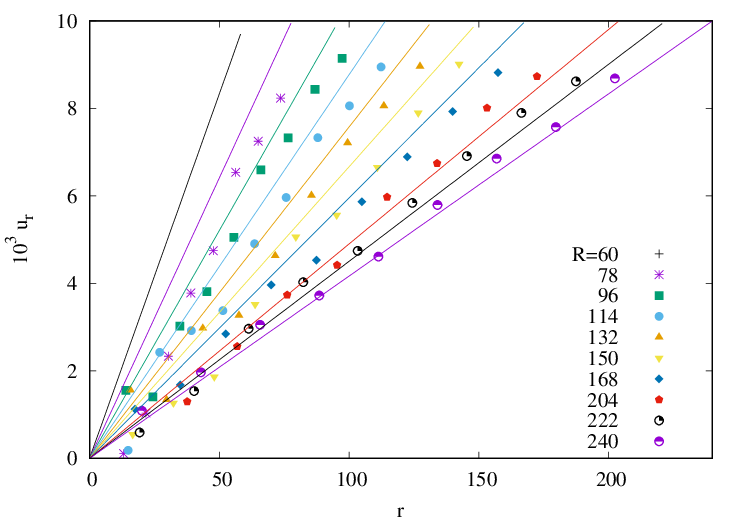}%
 \caption{Radial component of the velocity field \(u_r\) as a function of \(r\) for different system radii \(R\).}
 \label{fig:fig3}
\end{figure}

\begin{figure}[!h]
 \includegraphics[width=0.75\linewidth]{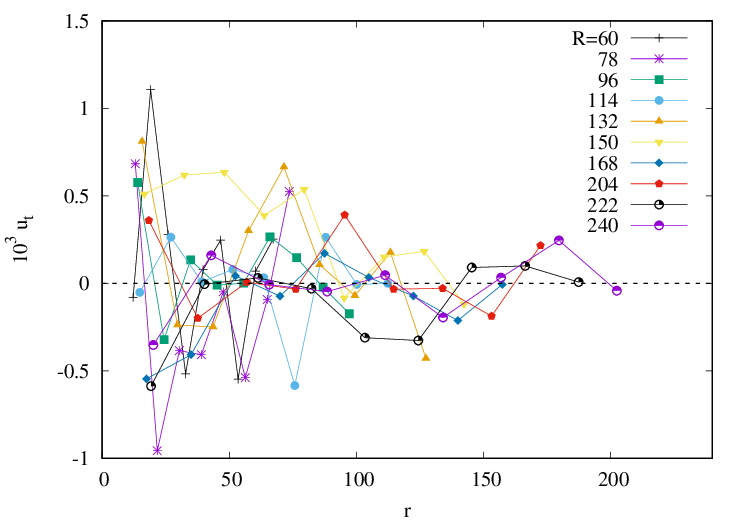}%
 \caption{Transversal component of the velocity field \(u_t\) as a function of \(r\) for different system radii \(R\).}
 \label{fig:fig4}
\end{figure}

\begin{figure}[!h]
 \includegraphics[width=0.75\linewidth]{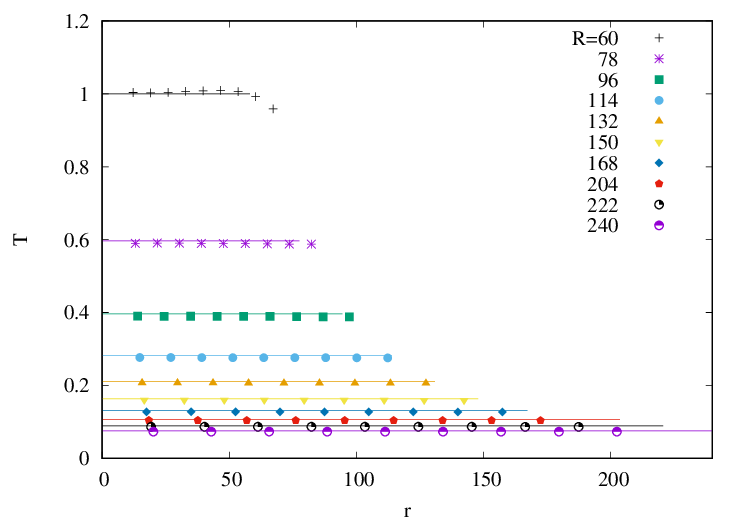}%
 \caption{Temperature \(T\) as a function of \(r\) for different system radii \(R\).}
 \label{fig:fig5}
\end{figure}

Apart from Fig.~\ref{m-fig:fig1} of the main text, Fig.~\ref{fig:fig2}--\ref{fig:fig5} provide further numerical results. 

\clearpage
\bibliography{eqkt}